%% file: main.tex
\documentclass[aps,prd,superscriptaddress,nofootinbib,floatfix,preprint]{revtex4-1}

\usepackage[dvipdfmx]{graphicx}   

\usepackage{comment}

\newcommand{\beq}{\begin{equation}}
\newcommand{\eeq}{\end{equation}}
\newcommand{\beqa}{\begin{eqnarray}}
\newcommand{\eeqa}{\end{eqnarray}}
\newcommand{\bpr}{\begin{problem}}
\newcommand{\epr}{\end{problem}}
\newcommand{\bcent}{\begin{center}}
\newcommand{\ecent}{\end{center}}
\newcommand{\bfig}{\begin{figure}}
\newcommand{\efig}{\end{figure}}
\newcommand{\bpc}{\begin{picture}}
\newcommand{\epc}{\end{picture}}
\newcommand{\barr}{\begin{array}}
\newcommand{\earr}{\end{array}}
\newcommand{\bitm}{\begin{itemize}}
\newcommand{\eitm}{\end{itemize}}
\newcommand{\bright}{\begin{flushright}}
\newcommand{\eright}{\end{flushright}}
\newcommand{\bminip}{\begin{minipage}}
\newcommand{\eminip}{\end{minipage}}
\newcommand{\btab}{\begin{tabular}}
\newcommand{\etab}{\end{tabular}}

\newcommand{\nnb}{\nonumber}

\newcommand{\hiroshima}{Graduate School of Advanced Science and Engineering, Hiroshima University, Kagamiyama, Higashi-Hiroshima 739-8526, Japan}

\newcommand{\icr}{Institute for Chemical Research, Kyoto University Uji, Kyoto 611-0011, Japan}

\newcommand{\kyoto}{Graduate School of Science, Kyoto University, Sakyouku, Kyoto 606-8502, Japan}

\begin{document}
\input{body.tex}

\input{ref.tex}
\end{document}

%% file: body.tex
\title{Extended search for sub-eV axion-like resonances via four-wave mixing with a quasi-parallel
laser collider in a high-quality vacuum system}

\author{Akihide Nobuhiro}\affiliation{\hiroshima}
\author{Yusuke Hirahara}\affiliation{\hiroshima}
\author{Kensuke Homma\footnote{corresponding author: khomma@hiroshima-u.ac.jp}}\affiliation{\hiroshima}
\author{Yuri Kirita}\affiliation{\hiroshima}
\author{Takaya Ozaki}\affiliation{\hiroshima}
\author{Yoshihide Nakamiya\footnote{currently IFIN-HH/ELI-NP, Romania}}\affiliation{\icr}
\author{Masaki Hashida}\affiliation{\icr}\affiliation{\kyoto}
\author{Shunsuke Inoue}\affiliation{\icr}\affiliation{\kyoto}
\author{Shuji Sakabe}\affiliation{\icr}\affiliation{\kyoto}

\date{\today}

\begin{abstract}
Resonance states of axion-like particles were searched for
via four-wave mixing by focusing two-color pulsed lasers into a quasi-vacuum.
A quasi-parallel collision system that allows probing of the sub-eV mass range was 
realized by focusing the combined laser fields with an off-axis parabolic mirror. 
A 0.10~mJ/34~fs Ti:Sapphire laser pulse and a 0.14~mJ/9~ns Nd:YAG laser pulse were 
spatiotemporally synchronized by sharing a common optical axis and focused into the vacuum system. 
No significant four-wave mixing signal was observed at the vacuum pressure of $3.7 \times 10^{-5}$~Pa
, thereby providing upper bounds on the coupling-mass relation 
by assuming exchanges of scalar and pseudoscalar fields 
at a 95~\% confidence level in the mass range below 0.21~eV.
For this search, the experimental setup was substantially upgraded so that optical
components are compatible with the requirements of the high-quality vacuum system, 
hence enabling the pulse power to be increased. With the increased pulse power,
a new kind of pressure-dependent background photons emerged in addition to the known atomic four-wave mixing 
process. This paper shows the pressure dependence of these background photons 
and how to handle them in the search.
\end{abstract}

\maketitle

\section{Introduction}
The spontaneous breaking of a global symmetry accompanies
a massless Nambu--Goldstone boson (NGB)~\cite{NGB} .
In nature, however, such an NGB possesses a finite mass due to complicated quantum corrections.
The neutral pion is such a pseudo-NGB (pNGB) state, gaining mass because of
chiral symmetry breaking in quantum chromodynamics (QCD).
The concept of spontaneous symmetry breaking can be a robust guiding principle 
to understand dark matter and dark energy in the context of particle physics.
Indeed, several theoretical models predict low-mass pNGBs
such as the QCD axion~\cite{axion} for solving the strong CP problem, and 
pNGBs in the contexts of string theory~\cite{strings} and
unified inflation and dark matter~\cite{miracle}, which are commonly pseudoscalar fields.
As a scalar-field example, dilaton is predicted as a source of dark energy~\cite{dilaton}.
These weakly coupling pNGBs are known generically as axion-like particles (ALPs).
Because it is difficult to evaluate the physical masses of pNGBs theoretically, 
model-independent laboratory experiments are indispensable 
for determining the physical masses as comprehensively as possible at relatively low masses 
compared to those of high-energy charged-particle colliders.
Because a pNGB is a low-mass state and in principle unstable even if the lifetime is long, 
the resonance state can be produced directly by colliding massless particles such as photons
as long as the pNGB has coupling to photons.
In previous works, we have reported on the first search for scalar-type pNGBs with laser beams in a quasi-parallel collision system (QPS)~\cite{PTEP-EXP00},  and then the second search for sub-eV scalar and pseudoscalar fields in the QPS~\cite{PTEP-EXP01}. 

\begin{figure}[h]
\begin{center}
\includegraphics[scale=0.48]{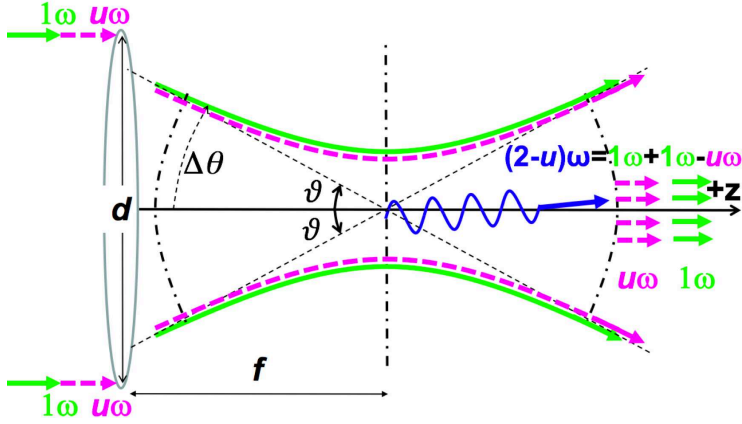}
\end{center}
\caption{A quasi-parallel collision system (QPS) combining two-color laser fields with beam diameter 
$d$ and focal length $f$, where the incident angle $\vartheta$ is subject to $0<\vartheta\leq \Delta\theta$, 
which is unavoidable because of the ambiguity of the wave vectors of the incident photons due to 
the nature of the focused lasers. This figure is reproduced from~\cite{PTEP-EXP01} with a slight modification.}
\label{Fig1}
\end{figure}

A QPS can be realized conceptually by 
focusing a laser beam with a focusing optical component as illustrated in Fig.~\ref{Fig1}.
The solid (green) laser field with photon energy $\omega$ is
focused, and the center of mass system (CMS) energy between a randomly selected 
photon pair within the focused field is expressed as  
\beq
E_{CMS}=2\omega\sin{\vartheta},
\eeq
where $\vartheta$ is half of the relative incident angle of the photon pair. 
By adjusting the beam diameter $d$ and focal length $f$, we can control
the range of possible incident angles within $\Delta\theta$, that is,
the accessible range of $E_{CMS}$ in a single focusing geometry.
Because $\vartheta$ can be close to zero, the QPS is 
in principle sensitive to pNGBs with almost zero mass. 
Consequently, the photon--photon scattering in the QPS
is accessible to the mass range $0< m < 2\omega\Delta\theta$,
although the sensitivity to a lower mass range is steeply suppressed.
Capturing a resonant pNGB within an $E_{CMS}$ uncertainty via an $s$-channel
exchange is the first key element of this method to increase the interaction rate. 
The second key element is in stimulation to guide the scattering
to a fixed final state. The dashed (magenta) laser beam with photon energy
$u\omega$ ($0<u<1$) indicates the inducing laser field to stimulate the interaction,
where the probability of generating a signal photon with energy $(2-u)\omega$ 
is enhanced via energy--momentum conservation in
$\omega+\omega \rightarrow pNGB \rightarrow u\omega + (2-u)\omega$
by the coherent nature of the co-propagating inducing laser.
The scattering probability increases in proportion 
to (i) the number of photons in the inducing laser
and (ii) the square of the number of photons in
the creation laser~\cite{DEptp,TajimaHomma,DEapb,DEptep}.
This process is kinematically similar to four-wave mixing 
in the context of atomic physics~\cite{FWM}.

Herein, we report the results of the third search by increasing the laser intensities with the upgraded experimental systems.
To increase the creation laser intensity in particular, we used a Ti:Sapphire-based laser, $T^6$-laser
at the Institute for Chemical Research at Kyoto University, which can produce a 10~TW pulse intensity
at maximum peak power at a repetition rate of 5~Hz. 
For this upgrade, main optical components were installed in the vacuum system which consists of 
transport chambers and an interaction chamber separately. We mostly respected construction of the high-quality vacuum
system,  especially for the interaction part, which required downsizing of the optical paths
and also the capability to manipulate optical components from outside these chambers.
The main purpose of the present search is therefore to show the extensibility of this searching method
in such a high-quality vacuum system, even if new types of background sources emerge 
as a result of increased laser intensities.

\section{Coupling-mass relation configured for the search}
In order to discuss the coupling of scalar ($\phi$) or pseudoscalar ($\sigma$) fields
to two photons, we introduce the following two effective Lagrangians: 
\beq
-L_{\phi} = gM^{-1}\frac{1}{4}F_{\mu\nu}F^{\mu\nu}\phi , \hspace{10pt} -L_{\sigma} = gM^{-1}\frac{1}{4}F_{\mu\nu}\tilde{F}^{\mu\nu}\sigma,
\label{eq_phisigma} 
\eeq
where 
$F^{\mu\nu}=\partial^{\mu}A^{\nu}-\partial^{\nu}A^{\mu}$ is the field strength tensor
and its dual $\tilde{F}^{\mu\nu}$ is defined as $\epsilon^{\mu\nu\alpha\beta}F_{\alpha\beta}$ with
the Levi-Civita symbol $\epsilon^{ijkl}$, 
and $g$ is a dimensionless constant while $M$ is a typical energy at which 
a relevant global symmetry is broken.

The scalar and pseudoscalar fields couple to the two photons differently depending
on the linear polarization states of the two photons.
In the photon-photon scattering process $p_1 + p_2 \rightarrow p_3 + p_4$ in four-momentum space,
when all photons are on an identical reaction plane, that is, in the case of the coplanar condition,
the distinction between scalar and pseudoscalar field exchanges becomes prominent as follows.
Denoting individual linear polarization states in parentheses,
the non-vanishing scattering amplitudes are limited to
\beqa
p_1(1) + p_2(1) \rightarrow p_3(2) + p_4(2) \nnb \\
p_1(1) + p_2(1) \rightarrow p_3(1) + p_4(1)
\eeqa
for the scalar field exchange and
\beqa
p_1(1) + p_2(2) \rightarrow p_3(1) + p_4(2) \nnb \\
p_1(1) + p_2(2) \rightarrow p_3(2) + p_4(1)
\eeqa
for the pseudoscalar field exchange, 
where swaps between (1) and (2) give the same scattering amplitudes, respectively.

\begin{figure}[!h]
\begin{center}
\includegraphics[scale=0.7]{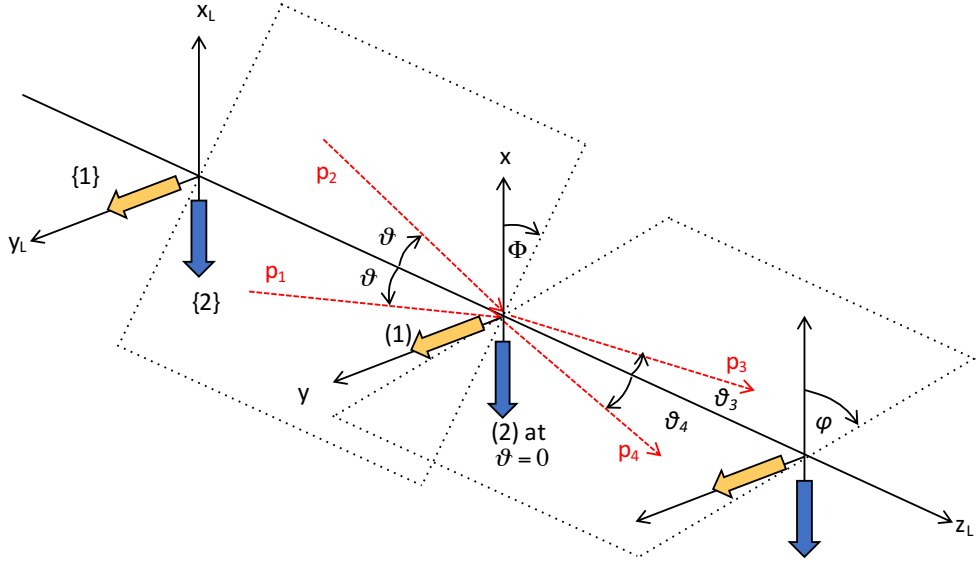}
\end{center}
\caption{
Definitions of linear polarization directions and rotation angles in focused collision geometry.
This figure is reproduced from \cite{PTEP-EXP01} with a slight modification.
}
\label{FigPOL}
\end{figure}

While this coplanar condition is always satisfied in CMS,
it is not true in the case of QPS as illustrated in Fig.\ref{FigPOL},
where the $\vec{p}_1-\vec{p}_2$ plane and the $\vec{p}_3-\vec{p}_4$ plane
may differ from the $x_L-z_L$ plane in the laboratory coordinates.
We define experimental linear polarization states \{1\} and \{2\}
by mapping them along $y_L$ and $x_L$ axes, respectively.
We then define rotation angles of these reaction planes as $\Phi$ and $\varphi$ for
initial and final state photon pairs, respectively, with respected to the $x_L-z_L$ plane.

In this search, indeed, we will assign the P-polarized state of the creation laser to \{1\} state
and the S-polarized state of the inducing laser to \{2\} state,
because the orthogonal combination of P- and S-polarization states is very effective
to suppress the known atomic four-wave mixing process in the residual gas~\cite{FWM,PTEP-EXP01}.

For a given set of fixed laboratory polarization directions \{1\} and \{2\},
we have introduced an axially asymmetric factor ${\cal F}_{abcd}$
with linear polarization states $ab$ and $cd$ for the initial and final state photon pairs,
respectively, by fixing $\Phi=0$ for scalar and pseudoscalar cases, respectively~\cite{DEptep}.
If there is no inducing field in the final state, $\vec{p}_3-\vec{p}_4$ planes rotate
around the $z_L$ axis symmetrically resulting in the axial symmetric factor of $2\pi$
due to the solid angle integration of signal photons.
However, since we induce $p_3$ photons as the signal by supplying the polarization-fixed
coherent $p_4$ field in the search, the $\vec{p}_3-\vec{p}_4$ plane rotation factor
deviates from $2\pi$ depending on types of exchanged fields. 
So the ${\cal F}$-factor corresponds to the replacement of the symmetric solid angle integral.

In addition we have further introduced an incident plane rotation factor ${\cal G}_{ab}$
in QPS~\cite{PTEP-EXP01},
because theoretically introduced $\vec{p}_1-\vec{p}_2$ planes are independent of
the experimentally introduced $x_L-z_L$ plane based on \{1\} and \{2\} states
of creation and inducing laser fields, respectively.
Thanks to this situation, even if the polarization direction of the creation laser field
is fixed at \{1\} state in the laboratory coordinates,
the searching system can also be sensitive to the pseudoscalar field exchange
through the rotation of the $\vec{p}_1-\vec{p}_2$ plane.
The ${\cal G}$-factor corresponds to an averaged correction factor
with respect to the case with $\Phi=0$ by taking possible rotation angles over $\Phi=0 \sim 2\pi$.

In this search we thus focus on the following scattering processes
in the same way as the previous publication~\cite{PTEP-EXP01}:
\beq
p_1(1) + p_2(1) \rightarrow p_3(2) + p_4(2)
\eeq
with the assumption of the scalar field exchange and
\beq
p_1(1) + p_2(2) \rightarrow p_3(1) + p_4(2)
\eeq
with that of the pseudoscalar field exchange.

For given parameters for the pulsed lasers and optical components, 
the coupling strength $g/M$ is related to the yield $\mathcal{Y}$ based on Eq.(\ref{eqYmod})
in Appendix herein, namely, the number of $p_3$ photons, 
via 
\beq
\frac{g}{M[\mbox{eV}]}= 2^{1/4} 8\pi^{2}
\sqrt{ \frac{ \mathcal{Y} \omega^3_c[\rm{eV}] }{ \left(\frac{\lambda_{c}}{c\tau_{c}}\right) 
\left(\frac{\tau_c}{\tau_i}\right)  \left(\frac{f}{d_c}\right)^{3}    
\frac{\pi}{4} \left(\frac{d_c}{d_i}\right)^2
\frac{(\overline{u}-\underline{u})^2}{\overline{u}\underline{u}}  
\mathcal{W} \mathcal{G}_{ab} \mathcal{F}_{abcd} C_{mb} {N_{c}}^2 N_{i}}}  m^{-5/2} [\rm{eV}],
\label{coupling}
\eeq
where 
the subscripts $c$ and $i$ denote the creation and inducing lasers, respectively. 
This relation is almost the same as the one used for the second search~\cite{PTEP-EXP01}
except for the condition that the beam diameters of the creation and inducing lasers are 
non-negligibly different in the present search, 
so the common diameter parameter for both beams cannot be assigned.
The relevant modifications are summarized in Appendix herein,
while all the other details are given in the appendices of the previous searches~\cite{PTEP-EXP00,PTEP-EXP01}.
The individual parameters are then summarized as follows:
$\omega_c$ is the incident photon energy of the creation laser,
$\lambda_{c,i}$ are the wavelengths of the two lasers, $\tau_{c,i}$ are the pulse durations, 
$f$ is the common focal length, $d_{c,i}$ are the beam diameters, 
$\mathcal{W}$ is the numerical factor relevant to the weighted integral of a Breit--Wigner resonance,
$\overline{u}$ and $\underline{u}$ are the upper and lower values, respectively, of $u$ determined by 
the spectrum width of $\omega_i$ based on $u \equiv \omega_i / \omega_c$, 
$C_{mb}$ is the combinatorial factor originating from 
selecting a pair of photons among multimode frequency states of the creation laser,
and $N_{c,i}$ are the mean numbers of photons per laser pulse.

\section{Experimental setup}
\begin{figure}[!h]
\begin{center}
\includegraphics[scale=1.2]{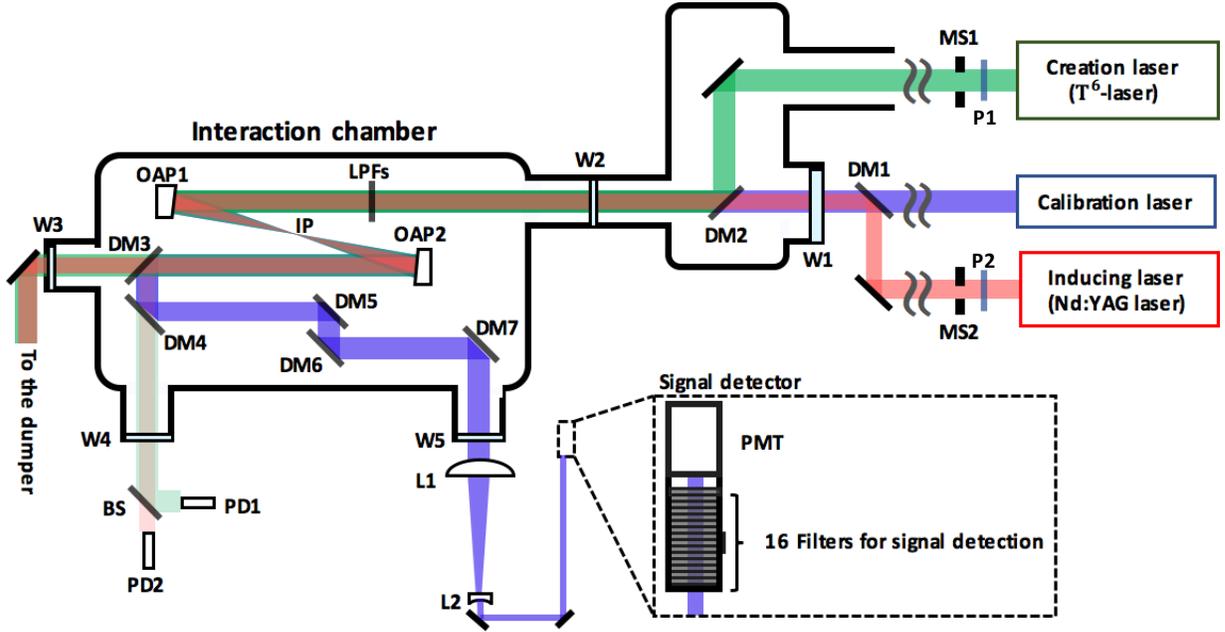}
\end{center}
\caption{Schematic of the searching system contained in a beam transport chamber
and an interaction chamber designed to achieve $10^{-8}$~Pa.}
\label{setup}
\end{figure}
The search was performed with the setup illustrated in Fig.~\ref{setup}.
A Ti:Sapphire pulsed laser with a central wavelength of 808~nm and a duration of 34~fs was used 
as the creation field, and an Nd:YAG pulsed laser with a central wavelength of 1064~nm and a duration 
of 9~ns was used as the inducing field. 
The linear polarization state of the creation laser was produced at P1 before the pulse compression.
P1 consisted of 30 synthetic quartz plates tilted at Brewster's angle to transmit
only p-polarized waves. The measured extinction ratio with the full energy shots was 
approximately 1000 (p-pol.) : 1 (s-pol.).
The linear polarization state of the inducing laser was produced at P2 
with a commercial polarization beam splitter with an extinction ratio 200 (p-pol) : 1 (s-pol.) 
and then the p-polarized waves were converted into s-polarized waves by a periscope.
The s-polarized waves can be further rotated by a half-wave plate for the systematic studies.
The creation and inducing lasers with beam diameters of 38~mm and 
16~mm, respectively, were combined by a dichroic mirror (DM2).
A 6-mm-thick window (W2) separated the vacuum systems between the interaction point (IP) 
and the upstream transport parts.
The combined beam was thus focused into the higher-quality vacuum part inside the interaction chamber
after penetrating through W2.
An off-axial parabolic mirror (OAP1) with a focal length of 279.1~mm was used for the focusing. 
Signal photons appeared only when spatiotemporal synchronization was achieved between the two pulsed beams at IP.
Background photons with the same wavelength as the signal ones generated 
along the optical path before OAP1 were removed by five long-pass filters (LPFs) 
that cut the signal wavelengths but transmitted the creation and inducing wavelengths
in advance of OAP1.

\begin{figure}[!h]
\begin{center}
\includegraphics[scale=0.78]{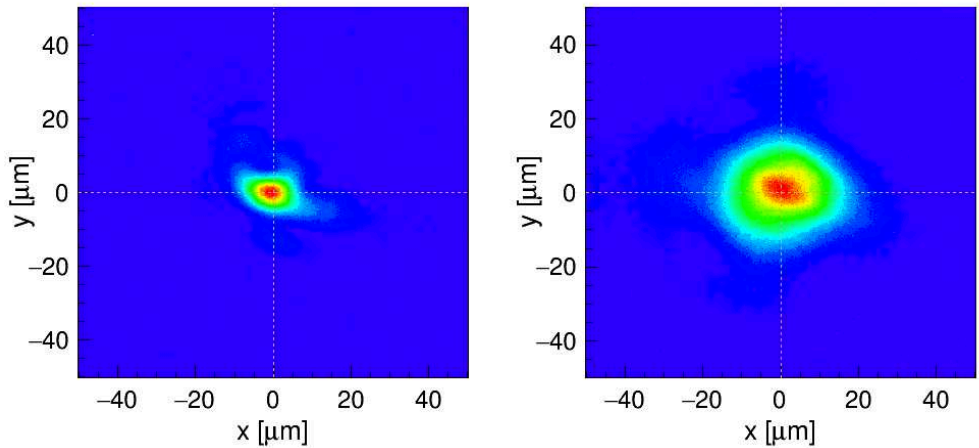}
\end{center}
\caption{Beam profiles of creation (left) and inducing (right) lasers at IP captured 
by a common CCD camera.}
\label{profile}
\end{figure}

Images of the beam profiles on the focal plane were transferred to a CCD camera
located outside the interaction chamber by an infinity-corrected optical system.
The camera recorded the number of beam photons per pixel with a spatial resolution of 0.3~$\mu$m. 
The physical image size was calibrated by transferring a mesh image located at IP 
whose physical size was known a~priori. 
The spatial synchronization of the two beams was ensured by adjusting optical components inside the transport
chambers for the creation laser and those outside the chambers for the inducing laser
so that the centers of the focal spots of the two beams coincided. 
The focal-spot images of the creation and inducing lasers used for the search are shown 
in Fig.~\ref{profile}. 
A beam diameter at IP is defined as the full width at half maximum (FWHM) of 
the 2-D Gaussian function fitted to a focal image.
The measured diameters were $\sigma_x = 10~\mu$m and $\sigma_y = 7~\mu$m 
in the horizontal $x$ and vertical $y$ directions for the creation laser,
while those of the inducing laser were $\sigma_x = 23~\mu$m and $\sigma_y = 19~\mu$m.

With the central wavelengths $\lambda_c$ and 
$\lambda_i$ for the creation and inducing lasers, respectively, 
the central wavelength $\lambda_s$ of the signal is defined via energy conservation as 
\beq 
\lambda_s=\left(\frac{2}{\lambda_c}-\frac{1}{\lambda_i}\right)^{-1} = 651\hspace{3pt} \mbox{nm},
\eeq
which is close to the 633~nm wavelength of a He:Ne laser.
Thus, the He:Ne laser was used to trace the signal trajectory down to the signal sensor
combined with the inducing laser by DM1 and then with the creation laser by DM2.
This calibration laser was used to align all the optical components inside the interaction chamber
and to evaluate the detector acceptance from IP to the sensor surface.

OAP1 and OAP2 were identical mirrors subtending IP at the same focal length.
The combined beam was focused by OAP1 and then collected by OAP2, which restored the plane waves.
The custom-made dichroic mirrors DM3 through DM7, 
each of which reflected 651~nm with 99\%  
while transmitting around 806~nm with 99\% and 1064~nm with 95~\%, 
discriminated the signal photons 
from the residual creation and inducing laser beams. 
The combined pulses that had passed through DM4 were measured 
by photodiodes (PD1, PD2) with $\sim40$~ps time-resolution to monitor the stability of the pulse energies 
and to adjust the time synchronization between the creation and inducing laser pulses.

As a signal detector, we used an R7400-01 single-photon-countable photomultiplier tube (PMT) 
manufactured by HAMAMATSU which has a falling time resolution of 0.75~ns.
In order for the detection system to be sensitive only to signal photons between 610 and 690~nm,
the PMT was installed after an LPF transmitting above 610~nm and 
three types of band-pass filters transmitting 570--800~nm, 500--930~nm, and 450--690~nm
to remove residual photons from the intense incident lasers 

\begin{figure}[!h]
\begin{center}
\includegraphics[scale=1.6]{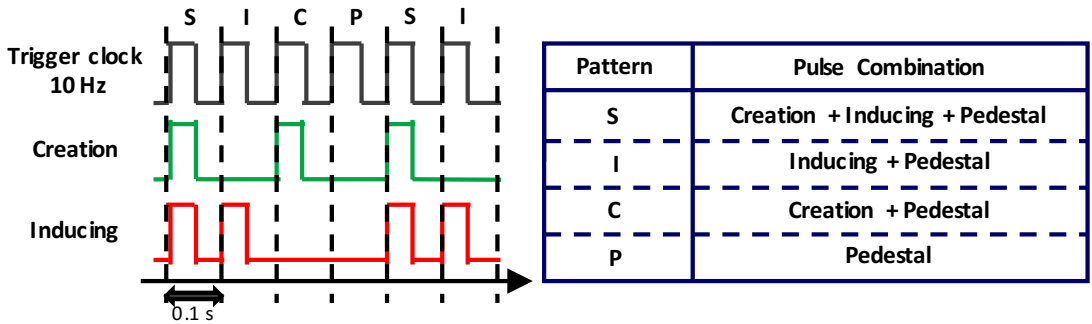}
\end{center}
\caption{Four consecutive trigger patterns.}
\label{Fig4}
\end{figure}

We acquired waveform data of analog currents from the PMT and the two PDs 
using a waveform digitizer triggered with a basic 10~Hz trigger to which both the creation and inducing
laser injections were synchronized as shown in Fig.~\ref{Fig4}.
Two mechanical shutters reduced the injection rate of the creation laser down to equi-interval 5~Hz and 
that of the inducing laser to non-equi-interval 5~Hz for waveform analysis as indicated in Fig.~\ref{Fig4}. 
The two beams were injected with different timing patterns so that the waveform data could be recorded 
with the following four consecutive trigger patterns with equal statistics:
(i) both beams were incident "S", (ii) only the inducing laser was incident "I", 
(iii) only the creation laser was incident "C", and (iv) neither beam was incident "P".

\section{Measurement of four-wave mixing process in residual gas}
\begin{figure}[!h]
\begin{center}
\includegraphics[scale=0.49]{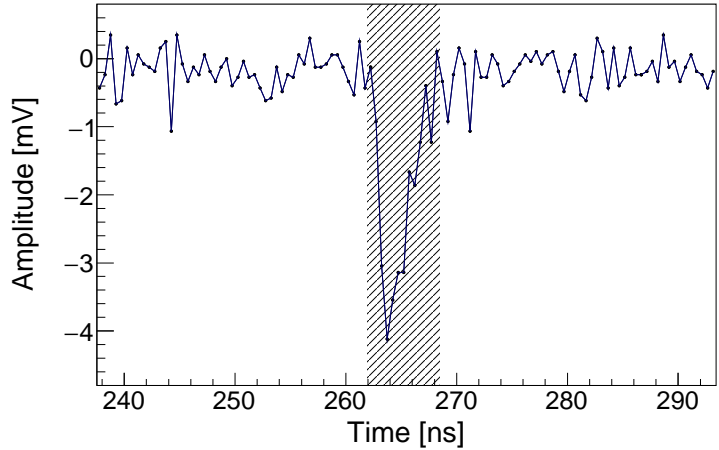}
\end{center}
\caption{Waveform data sample including a peak. The shaded area shows the integral range used to 
evaluate the charge sum of the peak structure.}
\label{wfm_sample}
\end{figure}

The observed number of photons was estimated by analyzing the waveform data from the PMT. 
An individual waveform consisted of 1000 sampling data points within a 500~ns time window. 
Figure~\ref{wfm_sample} shows an example of waveform data in which a peak is identified. 
 
Negative peaks were identified with a peak finder applying the same algorithm as 
that in the previous search~\cite{PTEP-EXP01}. 
We then calculated the charge sums in the peak structures.
The charge sums in the peak structures were converted to the number of photons by dividing 
by the measured single-photon equivalent charge.
%
\begin{figure}[!h]
\begin{center}
\includegraphics[scale=0.8]{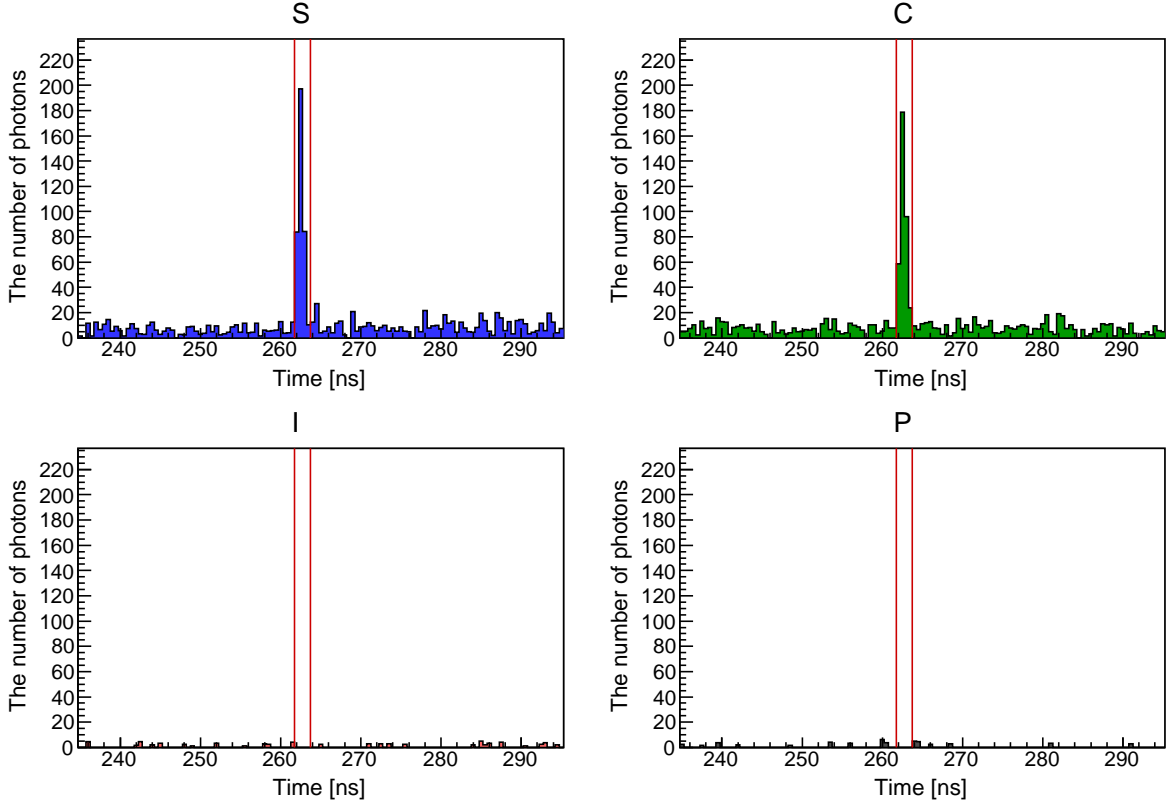}
\end{center}
\caption{Arrival time distributions of the number of observed photons in the case of the P-pol.(creation) + P-pol.(inducing) combination for individual trigger patterns at 10~Pa. 
Individual time windows in which signal photons are expected to be detected 
are indicated with the two red lines.}
\label{peak_atm}
\end{figure}
 
Four-wave mixing photons are expected to be caused by residual gases 
in the interaction chamber as a background source. 
However, this photon source can be used as a calibration source 
to assure the spatiotemporal synchronization of the creation and inducing laser pulses.
To quantify the number of background photons, we measured the pressure dependence of 
the number of four-wave mixing photons
when the linear polarization directions of the creation and inducing lasers 
were parallel to enhance the atomic four-wave mixing process. 
Figure~\ref{peak_atm} shows the arrival time distribution in units of the number of photons at 10~Pa 
for each of the four trigger patterns.
There is a peak structure in not only trigger pattern~S but also in pattern~C. 
The background photons in trigger pattern~C are expected to originate from plasma creation
by the ionization of the residual atoms when the high-intensity creation laser is focused at IP.
This phenomenon was not observed in the previous search~\cite{PTEP-EXP01} because of the 
lower intensity of the creation laser field.
\begin{figure}[!h]
\begin{center}
\includegraphics[scale=0.5]{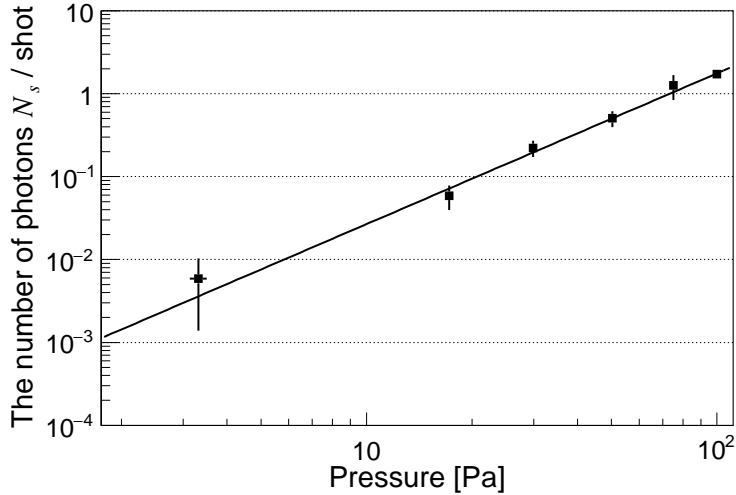}
\end{center}
\caption{Pressure dependence of the number of four-wave mixing photons per shot
from the residual atoms inside the interaction chamber
when the two beams with the same linear polarization directions,
P-pol.(creation) + P-pol.(inducing), are focused.}
\label{Fig7}
\end{figure}

The number of four-wave mixing photons, $N_{S}$, was then evaluated from 
\beq
N_{S} = n_{S} - n_{C} - n_{I} + n_{P},
\label{sub}
\eeq
where $n_{T}$ is the number of photons for a trigger pattern $T$
entering the time window 262.5--264~ns, which is consistent with the signal generation timing.
Figure~\ref{Fig7} shows the pressure dependence of the number of signal photons per shot, which 
was fitted with 
\beq\label{eq_scaling}
N_{S}/shot = a {\cal P}^{b},
\eeq
where $a$ and $b$ are fitting parameters and ${\cal P}$ is the pressure.
The error bars include the error propagation in the subtraction process based on 
statistical fluctuations of the reconstructed number of 
photons and uncertainties of focal-point stability during a run period.
The latter fluctuations were evaluated from overlaps between the two laser profiles 
measured by the common CCD camera which was sensitive to both wavelengths.
The overlap factor $O$ is defined as
\beq\label{eq_O0}
O \equiv \sum_{x}^{c} \sum_{y}^{c} N^2_c(x,y) N_i(x,y),
\eeq
where subscripts $c$ and $i$ indicate the creation and inducing lasers, respectively and 
$N(x,y)$ is the local intensity per CCD pixel of the monitor camera.
The summations were over the area framed by FWHM of the creation laser intensity profile.
Fluctuations of the overlap factors were then estimated as 
\beqa
\delta N_{S} = \left|N_{S} \frac{O_I - O_F}{O_I + O_F}\right|,
\label{eq_O}
\eeqa
where $O_{I,F}$ are the overlap factors at the beginning and end, respectively, of
a unit run period during 1200~s, and deviations with respect to 
the mean overlap factor $(O_I+O_F)/2$ were calculated.
The value of $b = 1.82 \pm 0.14$ from the fitting results  is close to the expected behavior $N_S \propto {\cal P}^2$ 
in atomic four-wave mixing processes~\cite{FWM, PTEP-EXP01}.

\begin{figure}[!h]
\begin{center}
\includegraphics[scale=0.55]{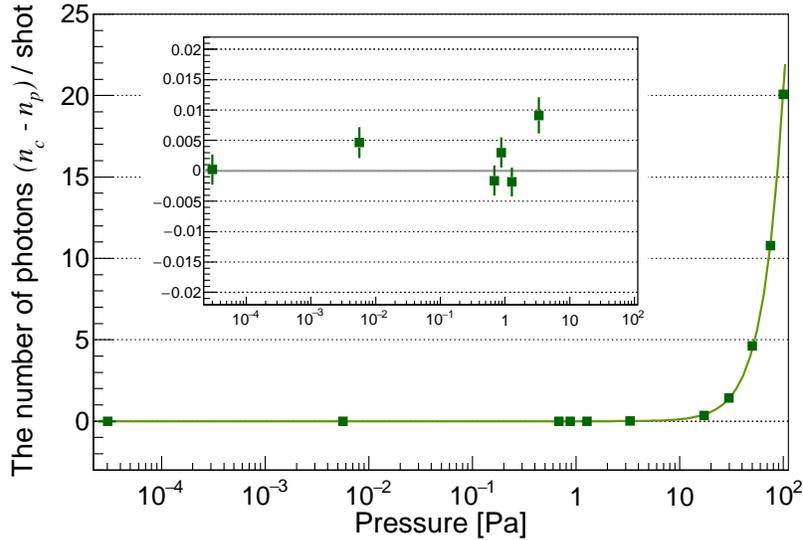}
\end{center}
\caption{Pressure dependence of the number of photons, $n_{c} - n_{p}$, per shot
when only linearly polarized creation laser pulses are focused in trigger pattern C. The errors are
statistical errors associated with subtracting side-band distributions
from the number of photons within the signal generation timing window.
Data points were fitted with $a{\cal P}^b + c$, 
resulting in $b = 2.22 \pm 0.01$ and $c=-6.58 \times 10^{-6} \pm 1.05 \times 10^{-3}$.
The inset figure zooms the constant points.
}
\label{Fig14}
\end{figure}

A notable difference from our previous publication~\cite{PTEP-EXP01} was that
signal-like photons were indeed produced even in trigger pattern~C.
Figure~\ref{Fig14} shows the number of photons, $n_{c} - n_{p}$, in trigger pattern~C within 
the signal timing window after subtracting the side-band background photons from the peak part
as a function of pressure. This indicates that $n_c - n_p$ scales with 
${\cal P}^{2.2}$ and reaches zero within the statistical uncertainty below 1~Pa.
Because $n_c - n_p$ is strongly pressure dependent, we conclude that this background source
is generated from IP but different from the conventional four-wave mixing process
because it does not require the inducing laser to produce the same frequency as that of four-wave mixing.
Creation of a plasma state at IP by the higher-power creation laser might 
explain the broad-band emissions, though this is not directly relevant to the present search
because we can subtract trigger pattern~C from pattern~S.

\begin{figure}[!h]
\begin{center}
\includegraphics[scale=0.5]{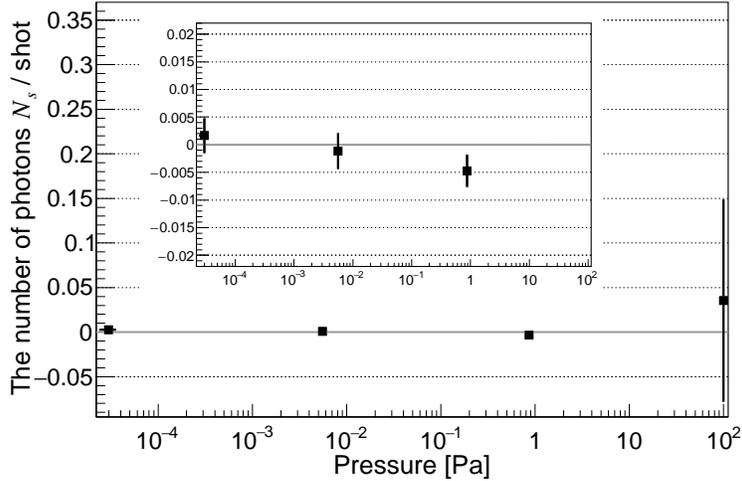}
\end{center}
\caption{Pressure independence of the number of four-wave mixing photons per shot
from the residual atoms inside the interaction chamber
when the two beams with the orthogonal linear polarization combination, P-pol.(creation)
+ S-pol.(inducing), are focused.}
\label{Fig8}
\end{figure}

\section{Search for four-wave mixing signals in vacuum}
We searched for scalar- and pseudoscalar-type resonance states 
at a vacuum pressure of $3.7 \times 10^{-5}$~Pa by combining P-pol. for the creation pulsed laser
and S-pol. for the inducing pulsed laser.

When the polarization directions of the two beams were orthogonal, P-pol.(creation) + S-pol.(inducing),
counting the number of four-wave mixing photons in the similar pressure range
to that of the P-pol.(creation) + P-pol.(inducing) case
was rather difficult because the number of photons in trigger pattern~C was considerable 
and the subtraction from trigger pattern~S was dominated by the statistical uncertainty.
However, we know that the parallel case dominates the orthogonal case
at a higher pressure range above $10^2$~Pa ~\cite{PTEP-EXP01}. 
Thus, the pressure dependence in Fig.~\ref{Fig7}
can give an upper limit on the number of photons from the atomic four-wave mixing process
even without the measured pressure dependence for the orthogonal case, which is 
the actual searching configuration as discussed in Section~II.
Figure~\ref{Fig8} shows the independence of the number of photons with respect to
pressure values in the case of the orthogonal polarization 
combination at a much lower pressure range compared to Fig.~\ref{Fig7}.
Although the subtraction error at around several Pa is much larger due to finite background photons
seen in trigger pattern C and S, the inset figure zooming the points below 1~Pa
indicates that the observed numbers of photons are consistent with null over five orders of
magnitude in the pressure values.

The upper limit on the expected number of background photons per shot (efficiency-uncorrected)
caused by residual gases at $3.7 \times 10^{-5}$~Pa was evaluated by extrapolating 
with Eq.~(\ref{eq_scaling}) based on the parallel polarization combination case as follows:

\beq\label{eq_Ngas}
N_{gas} / shot < 2.2 \times 10^{-11},
\label{total_n_gas}
\eeq
where the upper limit was estimated by taking the maximum value 
based on the fitting error on the parameter $b$.
Therefore, in the case of the orthogonal polarization combination,
the expected number of background photons caused by the residual atoms is below unity 
as long as the total statistic is below $10^{11}$ shots.

\begin{figure}[!h]
\begin{center}
\includegraphics[scale=0.8]{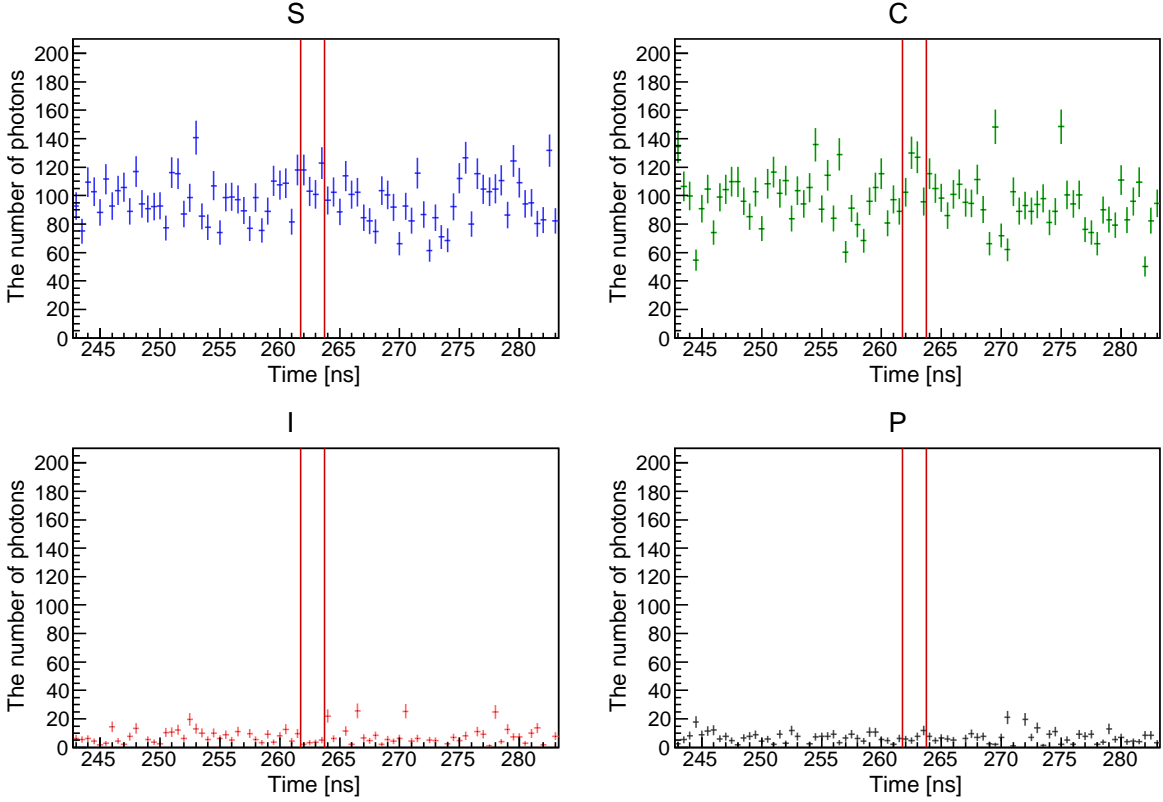}
\end{center}
\caption{Arrival time distributions of detected photons at $3.7\times10^{-5}$~Pa
for trigger patterns~S, C, I, and P with the P-pol.(creation) + S-pol.(inducing) combination.}
\label{Fig11}
\end{figure}

\begin{figure}[!h]
\begin{center}
\includegraphics[scale=0.5]{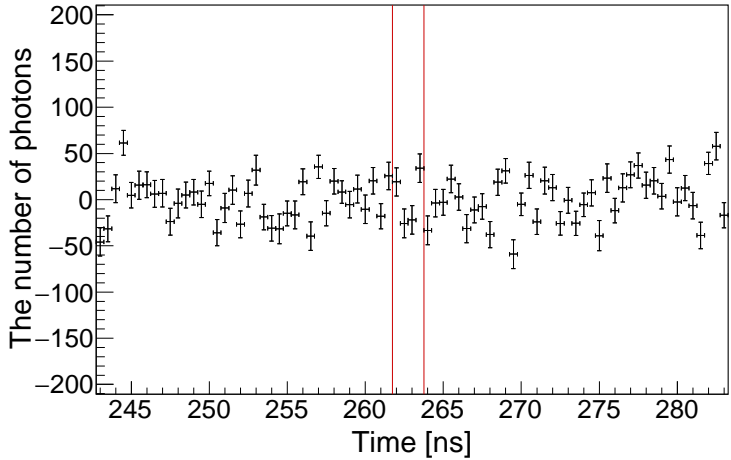}
\end{center}
\caption{Arrival time distribution of the number of photons obtained by applying Eq.~(\ref{sub})
to the other timing windows including that of the four-wave mixing signal enclosed by the two red lines
at $3.7\times10^{-5}$~Pa in the case of the orthogonal polarization combination,
P-pol.(creation) and S-pol.(inducing).}
\label{Fig12}
\end{figure}

Figure \ref{Fig11} shows the arrival time distributions of detected photons at $3.7 \times 10^{-5}$ Pa for
individual trigger patterns with the total number of shots in trigger pattern~S
defined as $W_S \equiv 4.2 \times 10^4$
in the case of the orthogonal polarization combination, P-pol.(creation) and S-pol.(inducing).
Figure \ref{Fig12} shows the arrival time distribution of the number of photons obtained 
by applying Eq.~(\ref{sub}) to the other timing windows including that of the four-wave mixing signal 
enclosed by the two red lines.
The total number of four-wave mixing photons within the signal generation timing
in the quasi-vacuum state was obtained as
\beq
N_{S} = 5.4 \pm 30.7(\rm{stat.}) \pm 25.1(\rm{syst.I}) \pm 10.1(\rm{syst.I\hspace{-1pt}I}) \pm 0.9(\rm{syst.I\hspace{-1pt}I\hspace{-1pt}I}).
\label{result}
\eeq
Systematic error~I is the uncertainty caused by the number of photons present 
outside the time window of the signal. This was estimated by measuring the root-mean-square of 
the number of photon-like signals excluding the signal window.
Systematic error~II originates from fluctuations of the overlap factors between
the focal spots of the creation and inducing lasers as discussed in Section~IV. 
We emphasize that Eq.(\ref{eq_O0}) includes fluctuations of the numbers of photons, $N_{c,i}$, observed
at the focal spots, namely, fluctuations of beam energies during a run period 
together with the position dependence, namely, pointing fluctuations.
The absolute values of laser pulse energies for the creation and inducing lasers were
$101 \pm 4$~$\mu$J and $300 \pm 5$~$\mu$J, respectively, at the beginning of the first run,
while they were  $98 \pm 4$~$\mu$J and $303 \pm 5$~$\mu$J, respectively,  at the end of the final run.
So the absolute pulse energies of the two lasers were quite stable over
the entire data taking period for 40~hours.
This overlap factor reflects the nature of four-wave mixing on the cubic intensity dependence.
Systematic error~III is an error caused by the threshold value used for the peak finding. 
This was calculated as two standard deviation assuming that the number of photon-like signals obtained 
by varying $V_{threshold}$ from $-1.2$ to $-1.4$~mV follows a uniform distribution.

\section{Excluded coupling-mass regions for scalar and pseudoscalar fields}
\begin{table}[h!]
\caption{Experimental parameters to set exclusion regions for the coupling-mass relations.
$\mathcal{G}^{sc}_{11}$ and $\mathcal{G}^{ps}_{12}$ are averaging factors
by the incident reaction plane rotation
for scalar and pseudoscalar field exchanges, respectively.
The evaluation of $\mathcal{G}$ is discussed in the appendix of the previous search~\cite{PTEP-EXP01}.
$\mathcal{F}^{sc}_{1122}$ and $\mathcal{F}^{ps}_{1212}$ are
axially asymmetric factors for scalar and pseudoscalar field exchanges, respectively,
as explained in the appendix of \cite{DEptep}.
Note that the effective $N_i$ is deduced from the overlap between the measured focal images 
of the inducing and creation lasers within the area framed by $\pm 3$ standard deviation
in the Gaussian-like intensity profile of the creation laser.
}
\begin{center}
\begin{tabular}{lr}  \\ \hline 
Parameter & Value \\ \hline
Total number of shots in trigger pattern~$S$, $W_S$   & $4.2 \times 10^4$ shots\\
Center of wavelength of creation laser $\lambda_c$   & 808 nm\\
Relative line width of creation laser ($\delta\omega_c/<\omega_c>)$ &  $2.0\times 10^{-2}$\\
Center of wavelength of inducing laser $\lambda_i$   & 1064 nm\\
Relative line width of inducing laser ($\delta\omega_{i}/<\omega_{i}>)$ &  $1.0\times 10^{-4}$\\
Duration time of creation laser pulse per injection $\tau_{c}$ & 34 fs \\
Duration time of inducing laser pulse per injection $\tau_{i}$ & 9 ns \\
Creation laser energy per $\tau_{c}$ & 101 $\pm$ 4 $\mu$J \\
Inducing laser energy per $\tau_{i} $ & 142 $\pm$ 2 $\mu$J \\
Focal length of off-axis parabolic mirror $f$ & 279.1~mm\\
Beam diameter of creation laser beam $d_{c}$ & 38 $\pm$ 0.8~mm\\
Beam diameter of inducing laser beam $d_{i}$ & 16 $\pm$ 0.3~mm\\
Upper mass range given by $\vartheta < \Delta\theta$ & 0.21 eV\\
$u=\omega_{i}/\omega_c$ & 0.76\\
Incident-plane-rotation averaging factor $\mathcal{G}_{ab}$
 &  $\mathcal{G}^{sc}_{11}$=19/32\\
 & $\mathcal{G}^{ps}_{12}$=1/2\\
Axially asymmetric factor $\mathcal{F}_{abcd}$
 & $\mathcal{F}^{sc}_{1122}$=19.4\\
 & $\mathcal{F}^{ps}_{1212}$=19.2\\
Combinatorial factor in luminosity $C_{mb}$  & 1/2\\
Single-photon detection efficiency $\epsilon_{D}$ & 1.4 $\pm$ 0.1 \% \\
Efficiency of optical path from IP to PMT $\epsilon_{opt}$ & 32.6 $\pm$ 4.0 \% \\
$\delta{N}_{s}$ & 40.9\\
\hline
\end{tabular}
\end{center}
\label{Tab1}
\end{table}

No statistically significant four-wave mixing photons in the quasi-vacuum state
were observed in this search from the result in (\ref{result}).
We thus set the exclusion regions on the coupling-mass relation by
assuming scalar and pseudoscalar fields based on the parameter values summarized in Tab.\ref{Tab1}.

First, the upper limit on the sensitive mass range is estimated as
\beq
m < 2\omega_c\sin\Delta\theta \sim 2\omega_c \frac{d_c}{2f} = 0.21 \mbox{~eV}
\eeq
based on values in Tab.\ref{Tab1}, where 
$\Delta\theta$ is defined by the focal length $f$ and beam diameter $d$ of the creation laser in Fig.~\ref{Fig1}
and $\vartheta$ varies from zero to $\Delta\theta$.

The efficiency-corrected number of four-wave mixing photons, $\mathcal{N}_{S}$, was evaluated 
from the following relation with the relevant experimental parameters: 
\beq 
\mathcal{N}_{S} = \frac{N_{S}}{\epsilon_{opt}\epsilon_{D}},
\label{yeild}  
\eeq
where $\epsilon_{opt}$ is the acceptance factor for signal photons to propagate 
from IP through several optical components, and 
$\epsilon_{D}$ is the detection efficiency of the PMT due mainly to the quantum efficiency 
of the photocathode when a single photon enters. 
$\epsilon_{opt}$ includes the inefficiencies from IP down to the actual location
of the PMT, namely, those of OAP2, DM3-DM7, W5, L1-L2, and the transmittance of the 16 wave filters in front of the PMT.
This factor was evaluated by taking the ratio between the calibration beam energy
at the focal point and that at the detection point.
The energy ratio was measured based on the intensity distributions at the common camera.
In advance of the search, $\epsilon_{D}$ was measured with 532~nm laser pulses.
We corrected the difference of the quantum efficiencies between 532~nm 
and 651~nm based on the relative quantum efficiencies provided by the HAMAMATSU specification sheet.

\begin{figure}[!h]
\begin{center}
\includegraphics[scale=0.68]{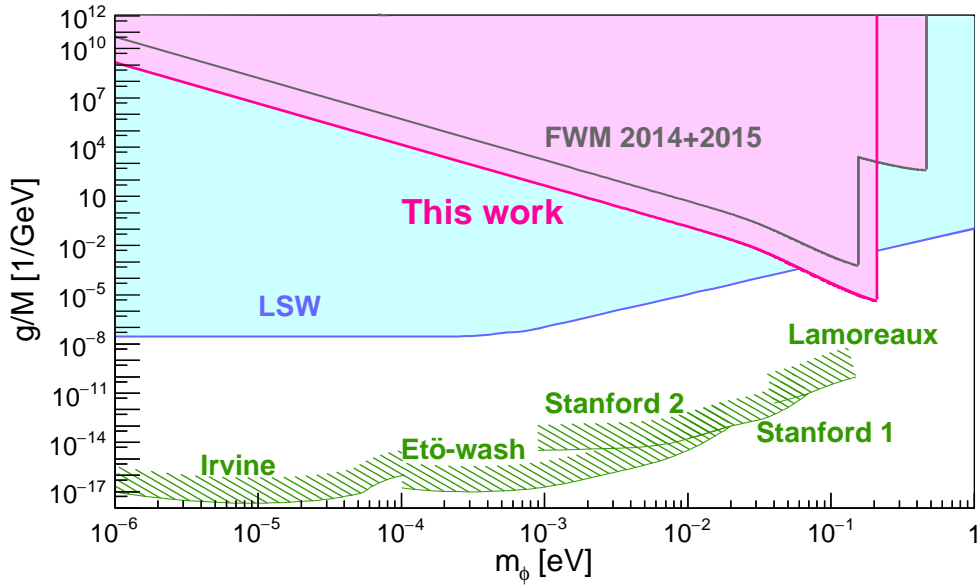}
\end{center}
\caption{Excluded regions in the coupling-mass relation for scalar-type fields ($\phi$).
The magenta area represents the excluded region by our method, "Four-Wave Mixing (FWM)". 
The thick magenta line is the new limit by this search,
while the gray line is the merged result from previous searches~\cite{PTEP-EXP00,PTEP-EXP01}.
The light blue area is the excluded region for scalar fields 
by the "Light Shining through a Wall (LSW)"  experiments (OSQAR~\cite{osqar} and ALPS~\cite{alps})
with simplification of the sine-function part to unity above $10^{-3}$~eV for drawing purpose.
The green shaded areas show regions excluded by non-Newtonian force searches 
by torsion-balance experiments "Irvine"~\cite{Irvine}, 
"Eto-wash"~\cite{Eto-wash}, "Stanford1"~\cite{Stanford1}, 
"Stanford2"~\cite{Stanford2} and Casimir force measurement "Lamoreaux"~\cite{Lamoreaux}.
}
\label{coupling_sc}
\end{figure}

\begin{figure}[!h]
\begin{center}
\includegraphics[scale=0.68]{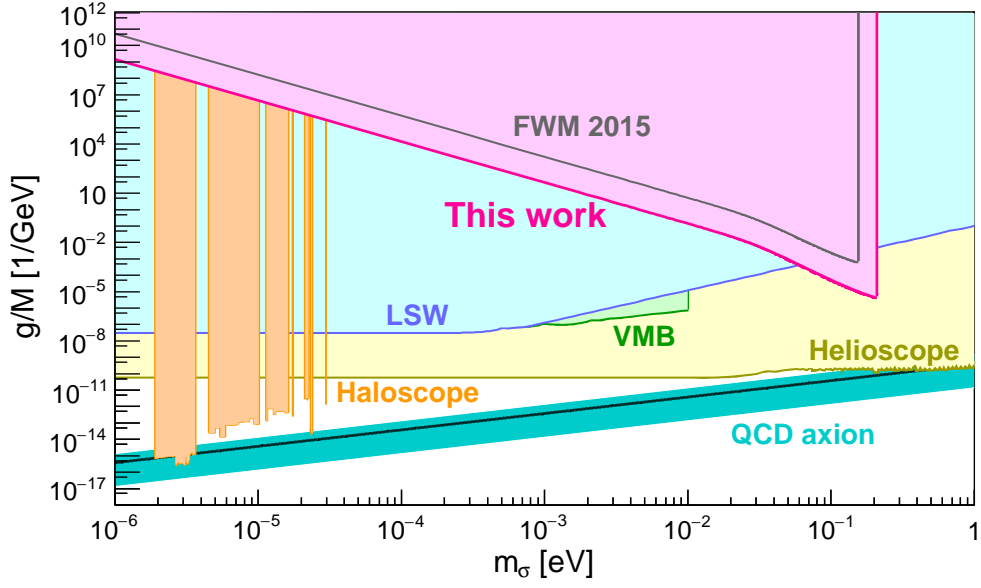}
\end{center}
\caption{
Excluded regions in the coupling-mass relation for pseudoscalar-type fields ($\sigma$).
This search based on FWM excludes the magenta area, while
the gray line is the limit by the previous search~\cite{PTEP-EXP01}.
The cyan band indicates the expected coupling-mass relation of the QCD axion predicted 
by the KSVZ model~\cite{KSVZ1,KSVZ2} with $|E/N-1.95|$ in the range 0.07--7.
The case of $E/N=0$ is also shown by the black solid line.
The light blue area indicates the excluded region by LSW experiments (OSQAR~\cite{osqar} and ALPS~\cite{alps}) 
for the pseudoscalar search.
The green area represents the result from the "Vacuum Magnetic Birefringence (VMB)" experiment (PVLAS~\cite{pv}). 
The yellow area shows the exclusion limit from the "Helioscope" experiment (CAST~\cite{cast}).  
The orange areas indicate excluded regions from "Haloscope" experiments (ADMX~\cite{admx}, RBF~\cite{rbf}, UF~\cite{uf} and HAYSTAC~\cite{haystac}). 
}
\label{coupling_ps}
\end{figure}

The upper limits on the coupling-mass relation at a 95~\% confidence level were obtained
based on the null hypothesis that the fluctuations of the number of signal photons follow
Gaussian distributions whose expectation value, $\mu$, is zero for the given total number of shots,
$W_S=4.2 \times 10^4$. 
This null hypothesis is justified. First, the expectation value based on the unique
standard model process, that is, the QED photon-photon scattering process is negligibly small
at $E_{cms} < 1$~eV~\cite{PTEPGG} even if we add the stimulation effect~\cite{PTEP17}
for the given shot statistics.
Second, as discussed with inequality Eq.~(\ref{eq_Ngas}), 
the expectation value due to atomic processes in the residual gas is well below unity.
Therefore, the background source is limited to only noises in the experimental environment.
The systematic error I is the dominant component of the systematic errors. 
This was estimated by measuring the root-mean-square of the number of photon-like signals 
excluding the signal window. In fact, the measured $N_s$ distribution excluding 
the signal time window follows almost a perfect Gaussian distribution around its central value, 
though the central value itself has no specific meaning with respect to the signal timing window. 
Since the distribution is symmetric around the central value, 
a Gaussian distribution can be supported from the measurement whatever the physics behind it is.
The Gaussian distribution is also deduced from the principle point of view.
What we measure is accidentally incident photon-like waveforms 
in arbitrary time window ($e.g.$ thermal noises, ambient photons from lasers and not from lasers, 
electric noises, and so forth) which should follow binomial distributions 
if the incident probabilities are individually constant over the measuring time. 
Although the incident probabilities are quite low, the expectation values of 
background photon-like signals per time bin over the data acquisition time are around 100 
as seen in trigger pattern S and C in Fig.\ref{Fig11}. 
Therefore, we may expect that the numbers of background photons per time bin can be approximated 
as Gaussian distributions because the expectation value is large enough. 
$N_s$ then eventually corresponds to subtractions between these Gaussian distributions. 
We thus assume that the Gaussian distribution is the most natural null hypothesis in our search.
In order to set a confidence level $1-\alpha$ to exclude the null hypothesis,
we assign the acceptance-uncorrected uncertainty $\delta N_{S}$ 
which was evaluated as the quadratic sum of statistical and systematic errors 
in Eq.~(\ref{result}) as the one standard deviation $\sigma$ to the following Gaussian kernel:
\beq
1-\alpha = \frac{1}{\sqrt{2\pi}\sigma}\int^{\mu+\delta}_{\mu-\delta}
e^{-(x-\mu)^2/(2\sigma^2)} dx = \mbox{erf}\left(\frac{\delta}{\sqrt 2 \sigma}\right),
\eeq
where $\mu=0$ and $N_{S}$ corresponds to the estimator $x$ in our case.
In order to give a confidence level of 95~\% in this analysis,
we apply $2 \alpha = 0.05$ with $\delta = 2.24 \sigma$ where
we set a one-sided upper limit by excluding above $x+\delta$~\cite{PDGstatistics}.
The upper limit of the signal yield to be compared to the theoretical calculation,
namely, $\mathcal{Y}$, was then evaluated as 
\beq
\mathcal{Y} = \frac{2.24\delta{N}_{S}}{\epsilon_{opt}\epsilon_{D}W_{S}}.
\eeq
Based on the coupling-mass relations in Eq.~(\ref{coupling}),
Figs.~\ref{coupling_sc} and \ref{coupling_ps} show the obtained exclusion limits
for scalar and pseudoscalar fields, respectively, at a 95~\% confidence level.

\section {Conclusions} 
We performed an extended search for scalar and pseudoscalar fields via four-wave mixing
by focusing two-color pulsed lasers, namely, 0.10~mJ/34~fs at 808~nm and 0.14~mJ/9~ns at 1064~nm.  
The observed number of four-wave mixing photons in the quasi-vacuum state at $3.7 \times 10^{-5}$~Pa was 
$5.4 \pm 30.7(\rm{stat.}) \pm 25.1(\rm{syst.I}) \pm 10.1(\rm{syst.I\hspace{-1pt}I}) \pm 0.9(\rm{syst.I\hspace{-1pt}I\hspace{-1pt}I})$.
We thus conclude that no significant four-wave mixing signal was observed in this search. 
The expected number of four-wave mixing photons from the residual gas is sufficiently small 
based on the upper limit from the measurement of the pressure dependence.  
With respect to an assumption that uncertainties are dominated by systematic fluctuations
around the zero expectation value following the Gaussian distribution,
we provided the upper limits on the coupling-mass relations for scalar and pseudoscalar fields 
at a 95~\% confidence level in the mass range below 0.21~eV, respectively.
The upper limits on the coupling at $m=0.21$~eV were obtained as 
$4.6 \times 10^{-6}$~GeV${}^{-1}$ and 
$5.1 \times 10^{-6}$~GeV${}^{-1}$ 
for scalar and pseudoscalar fields, respectively.

\section {Discussion and future prospect}
For this extended search, we upgraded the searching system so that it can store multiple 
dichroic mirrors to select feeble signal photons among a huge number of background laser fields
co-linearly propagating with the signal photons in the high-quality vacuum system.
We prioritized construction of the high-quality vacuum system over rapid increase of the creation laser intensity,
which is designed to achieve $10^{-8}$~Pa if a metal shield is installed for the lid contact, 
by introducing the entrance window (W2 in Fig.~\ref{setup})
in the interaction chamber to completely decouple the upstream transport vacuum system 
where the pressure is much higher. 
However, this window material indeed became an obstacle against drastically increasing
the laser intensity, because high-intensity laser fields must penetrate the window through which 
a new background source was created in addition to the residual atoms around IP. 
Therefore, the intensity improvement for the creation laser
from the previously published search~\cite{PTEP-EXP01} was rather moderate in the present search.
In this search, however, we have shown clearly that the null result can be
obtained with the current searching method, and we succeeded in measuring 
the pressure scaling of the atomic four-wave mixing process and also
another plasma-like phenomenon down to $\sim 1$~Pa.
For the next upgraded search, we thus plan to eliminate this window. Instead, 
a differential pumping system will be inserted between the transport vacuum system
and the interaction chamber.
Therefore, we may expect that we can drastically improve the sensitivity
with $\sim 10^3$ times higher creation laser intensity available at the $T^6$-laser system
in the upgraded system.

\section*{Acknowledgments}
We express deep gratitude to Yasunori Fujii, who passed away in July 2019.
This study was motivated by his outstanding work on the dilaton model.
K.~Homma acknowledges the support of the Collaborative Research
Program of the Institute for Chemical Research of 
Kyoto University (Grant Nos.\ 2018--83, 2019--72, and 2020--85)
and Grants-in-Aid for Scientific Research
Nos.\ 17H02897, 18H04354, and 19K21880 from the Ministry of Education, Culture, Sports, Science and Technology (MEXT) of Japan.

\newpage

\section*{Appendix: modifications of the coupling-mass relation due to the beam diameter difference}
We basically use the coupling-mass relation that was used for 
the previous searches~\cite{PTEP-EXP00,PTEP-EXP01}, where a common beam diameter 
was assumed for both the creation $(c)$ and inducing $(i)$ laser beams.
However, here we slightly modify the parametrization by taking different
beam diameters $d_c$ and $d_i$ into account.
With respect to equations for relevant parameter used in the Appendix of \cite{PTEP-EXP00}, 
we put superscript${~}^{*}$ for the modified parameter notations as follows.
 
For the density factor for the creation beam, namely, ${\mathcal D}_c$ in Eq.~(A29), we have 
\beqa\label{eq_Dc}
{\mathcal D}_c &=& \int^0_{-f/c} dt
\int^{\infty}_{-\infty} dx
\int^{\infty}_{-\infty} dy
\int^{\infty}_{-\infty} dz
{\rho_c}^2(x,y,z,t)
\nnb\\
&=&
\frac{N^2_c}{\sqrt{2\pi}} \frac{1}{\pi{w_0}^2} \frac{1}{c\tau_c}
\frac{z_R}{c} \left[ \tan^{-1} \left(\frac{f}{z_R}\right) \right]
\eeqa
$\rightarrow$
\beqa\label{eq_Dcmod}
{\mathcal D}^{*}_c &=& \int^0_{-Z_R/c} dt
\int^{\infty}_{-\infty} dx
\int^{\infty}_{-\infty} dy
\int^{\infty}_{-\infty} dz
{\rho_c}^2(x,y,z,t)
\nnb\\
&=&
\frac{N^2_c}{\sqrt{2\pi}} \frac{1}{\pi{w_0}^2} \frac{1}{c\tau_c}
\frac{z_R}{c} \left[ \frac{\pi}{4} \right],
\eeqa
where 
we reduce the range of time integration from $-f/c \sim 0$
to $-Z_R \sim 0$ with focal length $f$ and Rayleigh length $z_R$,
because the effect of the diameter mismatch on the overlap between the creation and the inducing lasers
is expected to be larger near the surface of the focusing mirror.
We thus restrict the signal generation range only in the vicinity of the focal point.

For the inducible acceptance factor ${\mathcal A}_i$ in Eq.~(A32), we have 
\beqa\label{eq_A4}
{\mathcal A}_i \sim
4\delta{\mathcal U}
\left( \frac{\vartheta_r}{\Delta\theta} \right)^2
\eeqa
$\rightarrow$
\beqa
{\mathcal A}^{*}_i \sim
4\delta{\mathcal U}
\left( \frac{\vartheta_r}{\Delta\theta_4} \right)^2
=
\left( \frac{\Delta\theta}{\Delta\theta_4} \right)^2
4\delta{\mathcal U} 
\left( \frac{\vartheta_r}{\Delta\theta} \right)^2
=
\left( \frac{d_c}{d_i} \right)^2
4\delta{\mathcal U} 
\left( \frac{\vartheta_r}{\Delta\theta} \right)^2,
\eeqa
where the approximation
$\Delta\theta_4 \sim \Delta\theta$ is no longer used 
because the beam diameters are not common while the focal length $f$ is common.

For the overall density factor
${\mathcal D}_{c+i}$ including the inducing laser effect in Eq.~(A34), we have 
\beqa\label{eq_Dciall}
{\mathcal D}_{c+i} &\sim&
C_{mb} \frac{N^2_c}{\sqrt{2\pi}} \frac{1}{\pi{w_0}^2} \frac{1}{c\tau_c}
\frac{z_R}{c}
\left[\tan^{-1}\left(\frac{f}{z_R}\right)\right]
4\delta{\mathcal U} \left(\frac{\vartheta_r}{\Delta\theta}\right)^2
\left(\frac{\tau_c}{\tau_i}\right) N_i
\eeqa
$\rightarrow$
\beqa\label{eq_Dciall}
{\mathcal D}^{*}_{c+i} &\sim&
C_{mb} \frac{N^2_c}{\sqrt{2\pi}} \frac{1}{\pi{w_0}^2} \frac{1}{c\tau_c}
\frac{z_R}{c}
\left[\frac{\pi}{4} \left( \frac{d_c}{d_i} \right)^2\right]
4\delta{\mathcal U} 
\left(\frac{\vartheta_r}{\Delta\theta}\right)^2
\left(\frac{\tau_c}{\tau_i}\right) N_i.
\eeqa

For the resultant signal yield ${\mathcal Y}$ in Eq.~(A35), we have 
\beqa\label{eqY}
{\mathcal Y} &=& {\mathcal D}_{c+i} \overline{\Sigma}
\nnb\\
&\sim&
\frac{1}{64\sqrt{2}\pi^4}
\left(\frac{\lambda_c}{c\tau_c}\right)
\left(\frac{\tau_c}{\tau_i}\right)
\left(\frac{f}{d}\right)^3
\left[\tan^{-1}\left(\frac{\pi d^2}{4\ f\lambda_c}\right)\right]
\frac{(\overline{u}-\underline{u})^2}{\overline{u}\underline{u}}
\left(\frac{gm{\mbox[\textrm{eV}]}}{M{\mbox[\textrm{eV}]}}\right)^2 
\nnb \\ 
&\times&
\left(\frac{m{\mbox[\textrm{eV}]}}{\omega{\mbox[\textrm{eV}]}}\right)^3
{\mathcal W}{\mathcal F_S} C_{mb} N^2_c N_i,
\eeqa
$\rightarrow$
\beqa\label{eqYmod}
{\mathcal Y} &=& {\mathcal D}^{*}_{c+i} \overline{\Sigma}
\nnb\\
&\sim&
\frac{1}{64\sqrt{2}\pi^4}
\left(\frac{\lambda_c}{c\tau_c}\right)
\left(\frac{\tau_c}{\tau_i}\right)
\left(\frac{f}{d_c}\right)^3
\left[\frac{\pi}{4}\left( \frac{d_c}{d_i} \right)^2\right]
\frac{(\overline{u}-\underline{u})^2}{\overline{u}\underline{u}}
\left(\frac{gm{\mbox[\textrm{eV}]}}{M{\mbox[\textrm{eV}]}}\right)^2 
\nnb \\ 
&\times&
\left(\frac{m{\mbox[\textrm{eV}]}}{\omega_c{\mbox[\textrm{eV}]}}\right)^3
{\mathcal W}{\cal G}_{ab}{\cal F}_{abcd} C_{mb} N^2_c N_i
\eeqa
with notation replacements: $d \rightarrow d_c$, $\omega \rightarrow \omega_c$ and 
${\cal F}_S \rightarrow {\cal G}_{ab}{\cal F}_{abcd}$ in this paper.